\documentclass[aps,preprint]{revtex4}%
\usepackage{amsfonts}
\usepackage{amsmath}
\usepackage{amssymb}
\usepackage{graphicx}%
\begin{document}
\title{Quantum superposition principle and generation of ultrashort optical pulses}
\author{Gevorg Muradyan}
\author{Atom Zh. Muradyan}
\affiliation{Department of Physics, Yerevan State Univeristy, 1 A. Manoogian Yerevan
375025, Armenia}


\begin{abstract}
We discuss the propagation of laser radiation through a medium of quantum
prepared $\Lambda$-type atoms in order to enhance the insight into the physics
of QSPT generator suggested in Phys. Rev. A \textbf{80}, 035801 (2009). We
obtain analytical results which \ give a qualitatively corerct description of
the outcoming series of ultrashort optical pulses and show that for the case
of alkali vapor medium QS-PT generation may be implemented under ordinary
experimental conditions.

\end{abstract}
\received[Received text]{date}

\revised[Revised text]{date}

\accepted[Accepted text]{date}

\published[Published text]{date}

\startpage{1}
\maketitle
\tableofcontents

\section{Introduction}

Atomic coherence is a widely accepted resource in atomic and laser physics for
modifying the response of coherently prepared atomic systems and thereby
canceling or enhancing the light absorption/reflection near atomic resonance
[1-4]. Quantum prepared medium in fact is essentially a new state of matter,
named in [5] phaseonium. The basic idea behind almost all such models is to
modify the emissive and absorptive profiles and make them non-reciprocal.
Among most important applications of quantum prepared medium, both in theory
and experimental science,one should list adiabatic population transfer [6-9],
amplification and lasing without inversion [10-13], sub-recoil laser cooling
[14-17], light storage and stopping [18-21], atomic spectroscopy [22-26], etc.

On the other hand, many problems in atomic physics, such as precision optical
frequency metrology, excitation spectroscopy and quantum state engineering
[27-30], require a train of high-energy and high-repetition optical pulses.
The most convenient method to accomplish this requirement today is the
acousto-optic modulator [31], but still new principles and schemes are
suggested in literature. A promising approach to this end is the usage of
quantum coherency created between atomic levels in both bare and dressed
approximations. For example, a considerable attention has been drawn to the
parametric beating of a weak probe field in the scheme of stimulated Raman
scattering [32,33]. In this paper we suggest a new step in this direction.

We consider propagation of an incident wave through a medium of $\Lambda$-type
three-level atoms with a preliminary superposition between the low lying
doublet states, a simplest construction for experimental realizations. We
carry a detailed analysis of how the superposition nature of atomic states
splits the incident monochromatic wave into a train of ultrashort (and high
power) pulses in a three-level medium. Such a qualitatively new and dramatic
effect of superposition principle has already been identified in our previous
papers [34,35]. However, it has proceeded from a two-level medium and two
(probe and pump) optical fields, thus somewhat disguising the inherent ability
of superposition principle in this problem. Simultaneously, the technical
realization of originally proposed scheme seems stringent as much as it is
difficult to onset the preliminary quantum superposition between the energy
levels in the optical range of Bohr frequencies. Presented in this paper
scheme clarifies the main idea of [34,35] and extends its formalism into much
headmost ideological and experimental conditions using $\Lambda$-type
interaction scheme and only \textit{one} travelling wave.

\section{Adiabatic states of $\Lambda$-type atom}

The basic setup of interaction used throughout this article is presented in
Fig. 1. The lower atomic states $1$ and $2$ are separated in energy by
$E_{a_{2}}-E_{a_{1}}=\hbar\Delta_{0}$, and are coupled with the upper state by
a plane optical field of frequency $\omega$:%
\begin{equation}
\overrightarrow{E}=\overrightarrow{E}_{0}(z,t)\exp(i\omega
t-ikz)+\overrightarrow{E}_{0}^{\ast}(z,t)\exp(-i\omega t+ikz), \tag{1}%
\label{1}%
\end{equation}
where $k=\omega/c$ is the vacuum wave number.%

%TCIMACRO{\FRAME{ftbpFU}{3in}{2.0003in}{0pt}{\Qcb{Energy-level diagram for the
%resonant single-photon process. States $\left\vert a1\right\rangle $ and
%$\left\vert a2\right\rangle $ are of like parity whereas the upper state
%$\left\vert b\right\rangle $ is of opposite parity. Energy distance
%$\hbar\Delta_{0}$ between the doublet states is much smaller than the photon
%energy $\hbar\omega$.}}{}{fig.1.eps}{\special{ language "Scientific Word";
%type "GRAPHIC";  maintain-aspect-ratio TRUE;  display "USEDEF";
%valid_file "F";  width 3in;  height 2.0003in;  depth 0pt;
%original-width 0pt;  original-height 0pt;  cropleft "0";  croptop "1";
%cropright "1";  cropbottom "0";
%filename 'Fig.1.eps';file-properties "XNPEU";}} }%
%BeginExpansion
\begin{figure}
[ptb]
\begin{center}
\includegraphics[
height=2.0003in,
width=3in
]%
{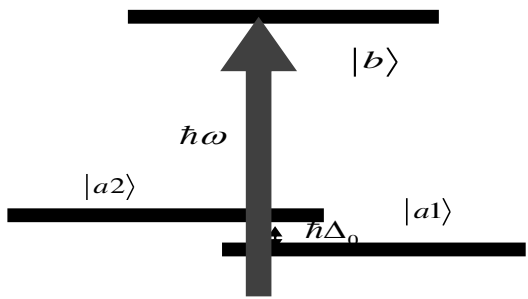}%
\caption{Energy-level diagram for the resonant single-photon process. States
$\left\vert a1\right\rangle $ and $\left\vert a2\right\rangle $ are of like
parity whereas the upper state $\left\vert b\right\rangle $ is of opposite
parity. Energy distance $\hbar\Delta_{0}$ between the doublet states is much
smaller than the photon energy $\hbar\omega$.}%
\end{center}
\end{figure}
%EndExpansion
The amplitude $\overrightarrow{E}_{0}(z,t)$ is assumed to vary slowly compared
to the rapidly oscillating exponential functions, i.e. $\left\vert
\partial\overrightarrow{E}_{0}(z,t)/\partial t\right\vert $ $<<\left\vert
\omega\overrightarrow{E}_{0}(z,t)\right\vert $ and $\left\vert \partial
\overrightarrow{E}_{0}(z,t)/\partial z\right\vert <<\left\vert
k\overrightarrow{E}_{0}(z,t)\right\vert $. (Later we will take it as a
constant when entering the gas medium, which is the most convenient form for
implication purposes). We also assume that the necessary conditions to
preserve the coherency are ensured. Dephasing from the upper state is
eliminated by assuming a far off-resonant interaction, while the relaxation or
dephasing between ground levels $1$ and $2$ lasts milliseconds and is
sufficient for the objectives of this paper. Thus we will discuss the
atom-radiation coupling staying in the limits of Schr\"{o}dinger equation%
\begin{equation}
i\hbar\frac{\partial\Psi(\overrightarrow{r},t)}{\partial t}=\left(
\widehat{H}_{0}-\widehat{\overrightarrow{d}}\overrightarrow{E}\right)
\Psi(\overrightarrow{r},t).\tag{2}\label{2}%
\end{equation}
The interaction is taken in electric dipole approximation, $\widehat{H}_{0}$
and $\widehat{\overrightarrow{d}}$ are free atom Hamiltonian and dipole moment
operator respectively, and the atomic motion isn't included in the discussion.

We look for the solution of Eq.2 in a form%
\begin{equation}
\Psi(\overrightarrow{r},t)=a_{1}(t)\psi_{1}(\overrightarrow{r})\exp
(-iE_{a_{1}}t/\hbar)+a_{2}(t)\psi_{2}(\overrightarrow{r})\exp(-iE_{a_{2}%
}t/\hbar)+b(t)\varphi(\overrightarrow{r})\exp(-iE_{b}t/\hbar)\text{,}
\tag{3}\label{3}%
\end{equation}
expanded over free atom doublet and excited state eigenstates $\psi
_{1,2}(\overrightarrow{r})$ and $\varphi(\overrightarrow{r})$ respectively.
$E_{a_{1,2}}$ and $E_{b}$ are corresponding energies. \ After substituting (
\ref{2}) into (\ref{3}) we apply the RWA neglecting terms with rapidly
oscillating factor $\exp(\pm i(\omega+\omega_{1,2})t)$ with $\omega
_{1,2}=E_{b}-E_{a_{1,2}}/\hbar$. We want to describe the phenomena under
conditions where the reactive part of the atom-field coupling is predominant
over the dissipative part so we proceed working in the range of resonance
detunings much larger than the Doppler and homogeneous widths of optical
transitions. This simultaneously gives the possibility of first order
adiabatic elimination procedure, algebraically presenting the excited state
amplitude through the lower lying ones:%
\begin{equation}
b(t)=-\frac{\overrightarrow{d}_{1}^{\ast}\overrightarrow{E}_{0}^{\ast}%
\exp(ikz)}{\hbar(\omega-\omega_{1})}\exp(-i(\omega-\omega_{1})t)a_{1}%
(t)-\frac{\overrightarrow{d}_{2}^{\ast}\overrightarrow{E}_{0}^{\ast}\exp
(ikz)}{\hbar(\omega-\omega_{2})}\exp(-i(\omega-\omega_{2})t)a_{2}(t).
\tag{4}\label{4}%
\end{equation}
Here $\overrightarrow{d}_{1}$ and $\overrightarrow{d}_{2}$ are $\ a1-b$ and
$a2-b$ optical transition matrix elements correspondingly.

For simplicity of later calculations we redefine the level 2 probability
amplitude%
\begin{equation}
A_{2}(t)=a_{2}(t)\exp(-i\Delta_{0}t),\text{ \ }A_{1}(t)=a_{1}(t),
\tag{5}\label{5}%
\end{equation}
and rescale the problem parameters to dimensionless ones:%
\begin{equation}
\tau\equiv\Delta_{0}t\text{, }\frac{\omega-\omega_{1}}{\Delta_{0}}\equiv
\omega-\omega_{1}\text{, }\frac{\omega-\omega_{2}}{\Delta_{0}}\equiv
\omega-\omega_{2}, \tag{6}\label{6}%
\end{equation}%
\begin{equation}
\xi_{1}\left(  z,\tau\right)  =\frac{\overrightarrow{d}_{1}\overrightarrow
{E}_{0}\left(  z,\tau\right)  }{\hbar\Delta_{0}}\text{, \ }\xi_{2}\left(
z,\tau\right)  =\frac{\overrightarrow{d}_{2}\overrightarrow{E}_{0}\left(
z,\tau\right)  }{\hbar\Delta_{0}}\text{.} \tag{7}\label{7}%
\end{equation}
The doublet amplitudes now will take a relatively simple form%
\begin{align}
i\frac{dA_{1}(\tau)}{d\tau}-\frac{\xi_{1}^{\ast}\left(  z,\tau\right)  \xi
_{1}\left(  z,\tau\right)  }{\omega-\omega_{1}}A_{1}(\tau)  &  =\frac{\xi
_{1}\left(  z,\tau\right)  \xi_{2}^{\ast}\left(  z,\tau\right)  }%
{\omega-\omega_{2}}A_{2}(\tau),\tag{8}\label{8}\\
i\frac{dA_{2}(\tau)}{d\tau}-A_{2}(\tau)-\frac{\xi_{2}^{\ast}\left(
z,\tau\right)  \xi_{2}\left(  z,\tau\right)  }{\omega-\omega_{2}}A_{2}(\tau)
&  =\frac{\xi_{2}\left(  z,\tau\right)  \xi_{1}^{\ast}\left(  z,\tau\right)
}{\omega-\omega_{1}}A_{1}(\tau)\text{.}\nonumber
\end{align}
They take up stationary solutions%
\begin{equation}
A_{1}(\tau)=A_{1}(0)\exp(-i\lambda_{1,2}\tau)\text{, \ }A_{2}(\tau
)=A_{2}(0)\exp(-i\lambda_{1,2}\tau) \tag{9}\label{9}%
\end{equation}
with characteristic values%
\begin{equation}
\lambda_{1}=\frac{1}{2}\left(  p+\sqrt{p^{2}-4q}\right)  \text{, \ \ }%
\lambda_{2}=\frac{1}{2}\left(  p-\sqrt{p^{2}-4q}\right)  \text{,}
\tag{10}\label{10}%
\end{equation}
where%
\begin{equation}
p=1+\frac{\xi_{1}^{\ast}\xi_{1}}{\omega-\omega_{1}}+\frac{\xi_{2}^{\ast}%
\xi_{2}}{\omega-\omega_{2}}\text{, \ \ \ }q=\frac{\xi_{1}^{\ast}\xi_{1}%
}{\omega-\omega_{1}}\text{.} \tag{11}\label{11}%
\end{equation}
These solutions should be normalized so that $\left\vert A_{1}(0)\right\vert
^{2}+$ $\left\vert A_{2}(0)\right\vert ^{2}=1$. \ Thus\ we obtained adiabatic
solutions, mapping the superposition of three bare states into a dressed
state. The solutions corresponding to the characteristic values $\lambda_{1}$
and $\lambda_{2}$ have the form
\begin{equation}
\Psi(t,\lambda_{1,2})=\left(
\begin{array}
[c]{c}%
A_{1}(0,\lambda_{1,2})\psi_{1}(\overrightarrow{r})\exp(-iE_{a_{1}}%
t/\hbar)+A_{2}(0,\lambda_{1,2})\psi_{2}(\overrightarrow{r})\exp(-iE_{a_{2}%
}t/\hbar+i\tau)\\
+b(0,\lambda_{1,2})\varphi(\overrightarrow{r})\exp(-iE_{b}t/\hbar)
\end{array}
\right)  \exp(-i\lambda_{1,2}t)\text{,} \tag{12}\label{12}%
\end{equation}
where%
\begin{align}
A_{1}(0,\lambda_{1,2})  &  =\frac{\xi_{2}^{\ast}\xi_{1}}{(\omega-\omega
_{2})(\lambda_{1,2}-q)\sqrt{1+\xi_{1}^{\ast}\xi_{1}\xi_{2}^{\ast}\xi
_{2}/(\omega-\omega_{2})^{2}(\lambda_{1,2}-q)^{2}}}\text{,} \tag{13}%
\label{13}\\
A_{2}(0,\lambda_{1,2})  &  =\frac{1}{\sqrt{1+\xi_{1}^{\ast}\xi_{1}\xi
_{2}^{\ast}\xi_{2}/(\omega-\omega_{2})^{2}(\lambda_{1,2}-q)^{2}}}%
\text{.}\nonumber
\end{align}

Both excited state amplitudes $b(0,\lambda_{1})$ and $b(0,\lambda_{2})$ are
determined from Eq.(\ref{4}).

\section{Light propagation in quantum prepared medium. Simple approximate
analytic solution}

Now we proceed to the discussion of the field propagation in the medium using
wave equation%
\begin{equation}
\left(  \frac{\partial^{2}}{\partial z^{2}}-\frac{1}{c^{2}}\frac{\partial^{2}%
}{\partial t^{2}}\right)  \overrightarrow{E}=\frac{4\pi\rho}{c^{2}}%
\frac{\partial^{2}}{\partial t^{2}}\left\langle \Psi(t)\right\vert
\widehat{\overrightarrow{d}}\left\vert \Psi(t)\right\rangle \text{,}
\tag{14}\label{14}%
\end{equation}
where the quantity $\rho\partial\left\langle \Psi(t)\right\vert \widehat
{\overrightarrow{d}}\left\vert \Psi(t)\right\rangle /\partial t$ stands for
the density of the bias electric current in dielectric gas medium, $\rho$ is
the number of atoms per unit volume and the brackets mean quantum mechanical
plus thermal state averaging. \ The form above in fact assumes that the gas
density is low enough and the temperature is high enough for quantum
mechanical collective effects to be irrelevant. We make the ansatz%
\begin{align}
\left\langle \Psi(t)\right\vert \widehat{\overrightarrow{d}}\left\vert
\Psi(t)\right\rangle  &  =\frac{1}{1+\exp\left(  -\frac{E_{a_{2}}-E_{a_{1}}%
}{k_{B}T}\right)  }\left\langle \alpha\Psi(t,\lambda_{1})+\beta\Psi
(t,\lambda_{2})\right\vert \widehat{\overrightarrow{d}}\left\vert \alpha
\Psi(t,\lambda_{1})+\beta\Psi(t,\lambda_{2})\right\rangle + \tag{15}%
\label{15}\\
&  \frac{1}{1+\exp\left(  \frac{E_{a_{2}}-E_{a_{1}}}{k_{B}T}\right)
}\left\langle \overline{\alpha}\Psi(t,\lambda_{1})+\overline{\beta}%
\Psi(t,\lambda_{2})\right\vert \widehat{\overrightarrow{d}}\left\vert
\overline{\alpha}\Psi(t,\lambda_{1})+\overline{\beta}\Psi(t,\lambda
_{2})\right\rangle \text{,}\nonumber
\end{align}
which assumes independent formation of superposition states from the thermally
populated levels $1$ and $2$. The coefficients $(\alpha,$ $\beta)$ and
$(\overline{\alpha},$ $\overline{\beta})$ are the probability amplitudes
corresponding to each channel.

Standard transformations in frame of slowly varying amplitudes approximation
bring us to a reduced form of wave equation (\ref{14}), which for a better
insight of the field dynamics we write in the following symbolic form:%
\begin{equation}
\left(  \frac{\partial}{\partial z}-\frac{1}{c}\frac{\partial}{\partial
t}\right)  \eta\left(  z,\tau\right)  =i\left(  F\left(  \left\vert
\eta\right\vert ^{2}\right)  +G_{C}\left(  \left\vert \eta\right\vert
^{2}\right)  \cos((\lambda_{2}-\lambda_{1})\tau)+G_{S}\left(  \left\vert
\eta\right\vert ^{2}\right)  \sin((\lambda_{2}-\lambda_{1})\tau)\right)
\eta\left(  z,\tau\right)  \text{.} \tag{16}\label{16}%
\end{equation}
The wave amplitude $\eta\left(  z,\tau\right)  $ is defined in a
dimensionless, symmetric with respect to both $1-3$ and $2-3$ transitions form%
\begin{equation}
\eta\left(  z,\tau\right)  =\frac{2d_{1}d_{2}E_{0}\left(  z,\tau\right)
}{(d_{1}+d_{2})\hbar\Delta_{0}}\text{.} \tag{17}\label{17}%
\end{equation}
The right-hand side expression in the brackets conditions the response of
medium to the coupling field and the nature of the field evolution. Note that
there is nothing surprising in the fact that the response of medium in
superposition state is time dependent, as is seen from Eq.(16), \ which leads
to phase modulation and relevant phenomena. The \textquotedblleft
hidden\textquotedblright\ from the first glance and crucial point here is the
existence of an imaginary part in the response function for this
non-dissipative medium under consideration. Periodic in time imaginary part
acknowledges the quantum superposition nature of atomic state prior to the
atom-field interaction and results in sequential amplification and suppressing
of the field intensity during the propagation. And this is the reason (or
origin) of the birth of series of short or ultrashort optical pulses, which we
term as quantum superposition pulse train (QS-PT) generator.

We will not discuss in this paper the questions connected with phase
modulation and proceed to a solution of the problem in terms of the field
intensity $\left\vert \eta\left(  z,\tau\right)  \right\vert ^{2}$. The wave
equation for it follows from Eq.(16):%
\begin{align}
\frac{\partial}{\partial\widetilde{\zeta}}\left\vert \eta\left(
\widetilde{\zeta},\widetilde{\tau}\right)  \right\vert ^{2} &  =i\frac
{2\pi\rho\omega d_{1}d_{2}}{\hbar\Delta_{0}^{2}}\left(  \frac{1}{\omega
-\omega_{2}}-\frac{1}{\omega-\omega_{1}}\right)  \times\nonumber\\
&  \left(
\begin{array}
[c]{c}%
\left(  \frac{\alpha^{\ast}\beta}{1+\exp\left(  -\frac{\hbar\Delta_{0}}%
{k_{B}T}\right)  }+\frac{\overline{\alpha}^{\ast}\overline{\beta}}%
{1+\exp\left(  \frac{\hbar\Delta_{0}}{k_{B}T}\right)  }\right)  \exp\left(
-i\left(  \lambda_{2}-\lambda_{1}\right)  \left(  \widetilde{\zeta}%
+\widetilde{\tau}\right)  \right)  -\\
\left(  \frac{\alpha\beta^{\ast}}{1+\exp\left(  -\frac{\hbar\Delta_{0}}%
{k_{B}T}\right)  }+\frac{\overline{\alpha}\overline{\beta}^{\ast}}%
{1+\exp\left(  \frac{\hbar\Delta_{0}}{k_{B}T}\right)  }\right)  \exp\left(
i\left(  \lambda_{2}-\lambda_{1}\right)  \left(  \widetilde{\zeta}%
+\widetilde{\tau}\right)  \right)
\end{array}
\right)  \times\tag{18}\label{18}\\
&  \left(  A_{2}\left(  0,\lambda_{1}\right)  A_{1}\left(  0,\lambda
_{2}\right)  -A_{1}\left(  0,\lambda_{1}\right)  A_{2}\left(  0,\lambda
_{2}\right)  \right)  \left\vert \eta\left(  \widetilde{\zeta},\widetilde
{\tau}\right)  \right\vert ^{2}\text{,}\nonumber
\end{align}
with $\widetilde{\zeta}=\Delta_{0}z/c$, $\widetilde{\tau}=\Delta_{0}(t-z/c)$.
\ Eq.(18) is the main result of this paper and the basis of later discussion.
We should note that in general it is a nonlinear equation and nonlinearity
enters through the intensity-dependence of characteristic values $\lambda_{1}$
and $\lambda_{2}$. As we can see they appear in the equation in two forms:
$A_{2}\left(  0,\lambda_{1}\right)  A_{1}\left(  0,\lambda_{2}\right)
-A_{1}\left(  0,\lambda_{1}\right)  A_{2}\left(  0,\lambda_{2}\right)  $ and
$\lambda_{2}-\lambda_{1}$. Numerical calculations, see Fig.2, show that the
first one of them, denoted $\left[  A_{2},A_{1}\right]  $ with high accuracy
may be treated as not depending on wave intensity. \ The difference
$\lambda_{2}-\lambda_{1}$ may also be viewed as as intensity independent, but
only when $\eta\left(  z,\tau\right)  <<1$ and equals to unity ($\lambda
_{2}-\lambda_{1}=1$). For this intensity range the problem becomes to a great
extend exactly soluble and has the following general solution form :%
\begin{equation}
\left\vert \eta\left(  z,t\right)  \right\vert ^{2}=\left\vert \eta\left(
0,t\right)  \right\vert ^{2}\exp\left(
\begin{array}
[c]{c}%
\frac{2\pi\rho\omega d_{1}d_{2}}{\hbar\Delta_{0}^{2}}\left(  \frac{1}%
{\omega-\omega_{2}}-\frac{1}{\omega-\omega_{1}}\right)  \times\\
\left(  A_{2}\left(  0,\lambda_{1}\right)  A_{1}\left(  0,\lambda_{2}\right)
-A_{1}\left(  0,\lambda_{1}\right)  A_{2}\left(  0,\lambda_{2}\right)
\right)  \times\\
\left(  \frac{\alpha^{\ast}\beta}{1+\exp\left(  -\frac{\hbar\Delta_{0}}%
{k_{B}T}\right)  }+\frac{\overline{\alpha}^{\ast}\overline{\beta}}%
{1+\exp\left(  \frac{\hbar\Delta_{0}}{k_{B}T}\right)  }\right)  \times\\
\frac{1-\exp\left(  -i\left(  \lambda_{2}-\lambda_{1}\right)  \left(
\Delta_{0}z/c\right)  \right)  }{\lambda_{2}-\lambda_{1}}\exp\left(  -i\left(
\lambda_{2}-\lambda_{1}\right)  \Delta_{0}t)\right)  +c.c.
\end{array}
\right)  \text{,}\tag{19}\label{19}%
\end{equation}
where the characteristic values $\lambda_{1}$ and $\lambda_{2}$ are determined
for the given initial wave-front.%
%TCIMACRO{\FRAME{ftbpFU}{3.8354in}{2.3082in}{0pt}{\Qcb{Functional form
%$[A_{2},A_{1}]$ (see the text) as function of scaled field intensity $\eta$ .
%Atomic transition frequencies $\omega_{1}=287351\Delta_{0}$ and $\omega
%_{2}=287350\Delta_{0}$ imply sodium atom's spectrum (see paragraph III of the
%main text), and input laser frequency $\omega=287360\Delta_{0}$ is detuned
%from these an order of magnitude larger than the doublet splitting $\Delta
%_{0}=$ $\omega_{1}-\omega_{2}$. }}{}{fig.2.eps}%
%width 3.8354in;  height 2.3082in;  depth 0pt;  original-width 3.787in;
%original-height 2.2684in;  cropleft "0";  croptop "1";  cropright "1";
%cropbottom "0";  filename 'Fig.2.eps';file-properties "XNPEU";}} }%
%BeginExpansion
\begin{figure}
[ptb]
\begin{center}
\includegraphics[
height=2.3082in,
width=3.8354in
]%
{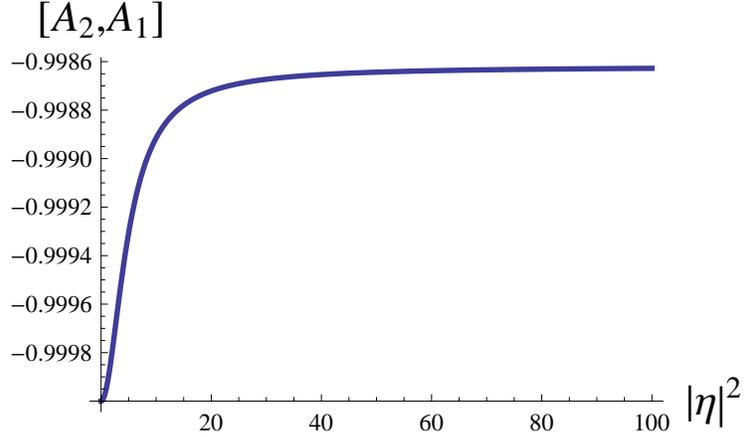}%
\caption{Functional form $[A_{2},A_{1}]$ (see the text) as function of scaled
field intensity $\eta$ . Atomic transition frequencies $\omega_{1}%
=287351\Delta_{0}$ and $\omega_{2}=287350\Delta_{0}$ imply sodium atom's
spectrum (see paragraph III of the main text), and input laser frequency
$\omega=287360\Delta_{0}$ is detuned from these an order of magnitude larger
than the doublet splitting $\Delta_{0}=$ $\omega_{1}-\omega_{2}$. }%
\end{center}
\end{figure}
%EndExpansion

This formula shows all the immense advantage of the scheme under
consideration: the output intensity is modulated at the frequency $\lambda
_{2}-\lambda_{1}$, which is the distance between the ground doublet levels.
This is a remarkable and a prominent performance index of the presented
scheme, since a very wide range of high stability repetition rates can be
attained for the pulse trains by simply choosing different atomic species with
doublet (or more) ground level structures.

The second prominent advantage of the presented scheme is the presence of a
prefactor $1/\Delta_{0}^{2}$ instead of \ $\frac{\Omega/\Delta}{\left(
1+(\Omega/\Delta)^{2}\right)  ^{2}}\frac{1}{\Delta^{2}}$ in the original
version [34], where $\Delta$ is detuning of the intense pump field relative to
the two-level optical transition. To neglect the spontaneous emission, one
usually has to work in the range where detunings are $\Delta>10^{11}Hz$, which
is at least one order of magnitude larger than $\Delta_{0}$ for alkaline metal
gases. At the same time in old version we are limited with $\Omega/\Delta=0.1$
upper value in the additional coefficient. So the present scheme possesses
much more freedom for such parameter values as gas density, and perhaps can be
implemented even on weak optical transitions.

Another preferable side of $\Lambda$-type interaction scheme is the
acquisition of superposition between the closely posed ground states, as it is
not a difficult task and may be implemented by means of magnetic fields as
well. In case of a preparative radiation field, one has $\alpha^{\ast}=\alpha
$, $\beta^{\ast}=-\beta$, $\overline{\alpha}^{\ast}=-\overline{\alpha}$ and
$\overline{\beta}^{\ast}=-\overline{\beta}$. The expression (19) becomes
somewhat simplified and the relative intensity $\left\vert \eta\left(
z,t\right)  \right\vert ^{2}/\left\vert \eta\left(  0,t\right)  \right\vert
^{2}$ (RI) takes the ultimate form%
\begin{equation}
\text{Relative intensity=}\exp\left(
\begin{array}
[c]{c}%
\frac{4\pi i\rho\omega d_{1}d_{2}}{\hbar\Delta_{0}^{2}}\left(  \frac{1}%
{\omega-\omega_{2}}-\frac{1}{\omega-\omega_{1}}\right)  \left(  \frac
{\alpha\beta}{1+\exp\left(  -\frac{\hbar\Delta_{0}}{k_{B}T}\right)  }%
+\frac{\overline{\alpha}\overline{\beta}}{1+\exp\left(  \frac{\hbar\Delta_{0}%
}{k_{B}T}\right)  }\right)  \times\\
\frac{A_{2}\left(  0,\lambda_{1}\right)  A_{1}\left(  0,\lambda_{2}\right)
-A_{1}\left(  0,\lambda_{1}\right)  A_{2}\left(  0,\lambda_{2}\right)
}{\lambda_{2}-\lambda_{1}}\left(  \sin(\left(  \lambda_{2}-\lambda_{1}\right)
(\tau-\Delta_{0}z/c))-\sin(\left(  \lambda_{2}-\lambda_{1}\right)
\tau)\right)
\end{array}
\right)  \tag{20}\label{20}%
\end{equation}

Expression in the exponent periodically oscillates as a function of time and
coordinate around the zero value. Therefore, in principle two types of field
envelope transformations are possible: propagation with a periodic
amplification and propagation with a periodic suppression. The answer to the
question which process will be realized for a given superposition state is
determined through the values of $\omega-\omega_{1}$ and $\omega-\omega_{2}$ differences.

\section{Numerical analysis of pulse train formation}

For quantitative analysis we discuss the case of $3S_{1/2}-$ $3P_{1/2}$
transition of $^{23}Na$. The hyperfine levels $F=1$ and $F=2$ of ground level
$3S_{1/2}$ constitute the doublet structure discussed. As the hyperfine
splitting in the excited state $3P_{1/2}$ is orders of magnitude \ smaller
than the $1771.6$ $MHz$ splitting of $3P_{1/2}$, we discuss the former as a
collection of degenerate sublevels of same energy. \ We are also allowed to
neglect all thermal transitions between ground levels $F=1$ and $F=2$ during
the interaction.

The incident wave is taken circularly polarized and this polarization is
preserved during propagation through the medium. Transition matrix elements
now should be calculated between the magnetic sublevels, and this should
include selection rules of the total moment $F$ and its projection $M$ on the
quantization axis aligned with the propagation direction. This, as usually, is
realized using formulas (see, for example [36])%
\begin{align*}
\left\langle F,M-1\right\vert \widehat{d}_{-}\left\vert F,M\right\rangle  &
=\sqrt{\frac{(F-M+1)(F+M)}{F(F+1)(2F+1)}}\left\langle F\right\vert \widehat
{d}\left\vert F\right\rangle \text{,}\\
\left\langle F,M-1\right\vert \widehat{d}_{-}\left\vert F-1,M\right\rangle  &
=\sqrt{\frac{(F-M+1)(F-M)}{F(2F-1)(2F+1)}}\left\langle F\right\vert
\widehat{d}\left\vert F-1\right\rangle \text{,}\\
\left\langle F-1,M-1\right\vert \widehat{d}_{-}\left\vert F,M\right\rangle  &
=\sqrt{\frac{(F+M+1)(F+M)}{F(2F-1)(2F+1)}}\left\langle F-1\right\vert
\widehat{d}\left\vert F\right\rangle ,\\
\left\langle F^{\prime},M^{\prime}\right\vert \widehat{d}_{+}\left\vert
F,M\right\rangle  &  =(\left\langle ^{\prime}F,M\right\vert \widehat{d}%
_{-}\left\vert F^{\prime},M\right\rangle )^{\ast}\text{,}%
\end{align*}
separating a projection-dependent prefactor from the reduced matrix element
$\left\langle F^{\prime}\right\vert \widehat{d}\left\vert F\right\rangle $.
\ Here $\widehat{d}_{\pm}=\widehat{d}_{x}\pm i\widehat{d}_{y}$ are circular
components of dipole moment operator. Finally it was convenient to use the
formulae%
\begin{equation}
\left\langle F^{\prime}\right\vert \widehat{d}\left\vert F\right\rangle
=\sqrt{\frac{3\hbar e^{2}}{2m\omega_{FF}}(2F+1)f(F\rightarrow F^{\prime}%
)},\tag{21}\label{21}%
\end{equation}
introducing the oscillator strength $f(F\rightarrow F^{\prime})$ instead of
reduced matrix element $\left\langle F^{\prime}\right\vert \widehat
{d}\left\vert F\right\rangle $.%
%TCIMACRO{\FRAME{ftbpFU}{3.9332in}{3.6512in}{0pt}{\Qcb{ Splitting of
%monochromatic incident wave intensity into a series of ultrashort pulses in a
%quantum prepared three-level atomic medium.  Here  $z\Delta_{0}/c=\pi$ ,
%$\rho=4\times10^{12}cm^{-3}$, $T=10^{-6}K$, $\alpha=\overline{\beta}=0.5$,
%$\beta=\overline{\alpha}=i\sqrt{0.75}$ . All the other parameters are the same
%as in Fig.2.}}{}{fig.3.eps}{\special{ language "Scientific Word";
%type "GRAPHIC";  maintain-aspect-ratio TRUE;  display "USEDEF";
%valid_file "F";  width 3.9332in;  height 3.6512in;  depth 0pt;
%original-width 3.8847in;  original-height 3.6045in;  cropleft "0";
%filename 'Fig.3.eps';file-properties "XNPEU";}} }%
%BeginExpansion
\begin{figure}
[ptb]
\begin{center}
\includegraphics[
height=3.6512in,
width=3.9332in
]%
{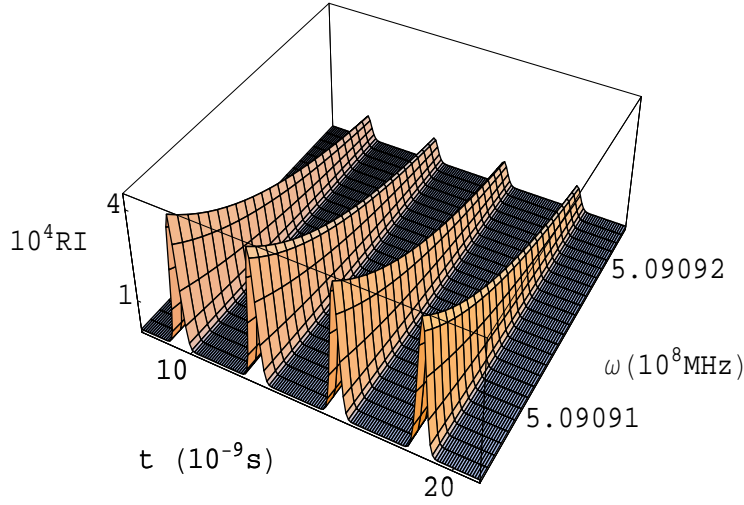}%
\caption{ Splitting of monochromatic incident wave intensity into a series of
ultrashort pulses in a quantum prepared three-level atomic medium.  Here
$z\Delta_{0}/c=\pi$ , $\rho=4\times10^{12}cm^{-3}$, $T=10^{-6}K$,
$\alpha=\overline{\beta}=0.5$, $\beta=\overline{\alpha}=i\sqrt{0.75}$ . All
the other parameters are the same as in Fig.2.}%
\end{center}
\end{figure}
%EndExpansion

The expression for the light field amplitude is of Eq.(19) form. The depth of
modulation increases with atom density and thereby decreases the duration of
each individual pulse in the sequence. Pulse duration may be significantly
controlled using resonance detuning too. Fig.3 graphically illustrates the
situation, plotting the envelope of the field intensity as a function of time
for various frequency detunings. Though the parameters are chosen for
illustrative purposes, they are realistic for ordinary laboratory conditions.
As might have been expected, pulse repetition rate is irrelevant to resonance
detuning and mimics the doublet splitting frequency. As to pulse duration, is
around\ hundred picoseconds for the chosen frequency region and concomitant
with power amplification. One will \ have the same picture for spatial
evolution too.

\section{Conclusions and outlook}

We showed that the gas of quantum prepared $\Lambda$-type atoms is very
efficient in splitting of monochromatic wave into a sequence of ultrashort
pulses, enlarging the spectrum of pulse generation mechanisms suggested or
implemented so far. We emphasize that the repetition rate of ultrashort pulses
does not depend neither on the light intensity nor its detuning from the
atomic resonance. The sole quantity determining the repetition rate is the
frequency separation between the ground doublet levels between which the
superposition has been preliminarily established. Duration of each pulse, vice
versa, is rigorously dependent on coupling parameters and effectively
controllable. These regularities accompanied by mild conditions for the
experimental realization presents the QS-PT generator as a viable alternative
to more conventional methods.

Finally we would like to note, that since the QS-PT generator is inevitably
grounded on the quantum superposition principle, the studied phenomenon of
abrupt shape-reformation of the classical electromagnetic field envelope may
be regarded as an example of quantum macroscopic effect.

\begin{acknowledgments}
This work was supported by the Armenian State Committee of Science.
\end{acknowledgments}

\end{document}